\documentstyle[12pt,twoside,fleqn,espcrc1,epsf]{article}
\newcommand{\be}{\begin{equation}}
\newcommand{\ee}{\end{equation}}
\newcommand{\bea}{\begin{eqnarray}}
\newcommand{\eea}{\end{eqnarray}}

\def\Journal#1#2#3#4{{#1} {\bf #2}, #3 (#4)}

\def\NPB{Nucl. Phys. {\bf B$\!\!$}}
\def\PLB{Phys. Lett. B}

\def\PRD{Phys. Rev. D$\!$}

\def\NPBPS{Nucl. Phys. B (Proc. Suppl.)}

\begin{document}

\title{Quark Confinement Physics in Quantum Chromodynamics}
\author{Y. Koma, H. Suganuma, K. Amemiya, M. Fukushima and H. Toki\\
\hspace{0.5cm}\\
Research Center for Nuclear Physics, Osaka University, 
Ibaraki, Osaka 567-0047, Japan}

\maketitle

\begin{abstract}
We study abelian dominance and monopole condensation for 
the quark confinement physics using the lattice QCD simulations 
in the MA gauge.
These phenomena are closely related to the
dual superconductor picture of the QCD vacuum, and 
enable us to construct the dual Ginzburg-Landau (DGL) theory
as an useful effective theory of nonperturbative QCD.
We then apply the DGL theory to the studies of the  
low-lying hadron structure and the scalar glueball properties.
\end{abstract}

\section{Introduction}

\par
Recent studies of the lattice QCD in the maximally 
abelian (MA) gauge suggest the remarkable properties of the QCD vacuum, 
such as abelian dominance\cite{lattice-abel} 
and monopole condensation\cite{lattice-mono},   
which provide the dual superconductor picture of the QCD vacuum 
as is described by the dual Ginzburg-Landau (DGL) theory\cite{sst}. 
In the MA gauge, QCD is reduced into an abelian gauge theory 
including color-magnetic monopoles.
According to the lattice QCD results, the nonperturbative quantities as 
the string tension and the chiral condensate are almost reproduced only by
the diagonal gluon part, while the off-diagonal gluon does not 
contribute to such the long-range physics, namely, abelian dominance.
Furthermore, the world-line of the color-magnetic monopole in the 
confinement phase appears as the global network, which indicates 
monopole condensation.
Then, the DGL theory can be constructed by extracting the diagonal gluon 
as the relevant degrees of freedom and taking into account monopole 
condensation.
Based on the DGL theory, the quark confinement is explained by
the flux-tube formation through the dual Meissner effect, and
chiral symmetry breaking is described as the function of monopole 
condensate\cite{sst}.

\par
In this paper, we focus such the dual superconductor picture of the 
QCD vacuum in the MA gauge, and confirm the connection between 
nonperturbative QCD and the DGL theory.
Then, we would like to apply the DGL theory to hadron physics, 
especially, to the analysis of the scalar glueball properties.

\section{Abelian dominance and monopole condensation in the MA gauge}

\par
Abelian dominance and monopole condensation in the MA gauge are 
the keywords to connect the QCD with the DGL theory, and 
the recent lattice QCD simulations show
the former on the string tension and the chiral condensate, 
and the latter as the large clustering of the monopole world-line.
In such situation, we still have interest in this subject,
since the physical essence of abelian dominance is not understood yet.
Furthermore, we must not jump to a conclusion that 
the global network of the monopole world-line 
is really 
the evidence of monopole condensation, which
also should be evaluated quantitatively.

\par
To answer these questions, 
we first study the gluon propagator in the MA gauge and 
evaluate the mass of the off-diagonal gluon field
using the SU(2) lattice QCD simulation\cite{amemiya}.
This study is based on the following idea.
If the off-diagonal gluon has a mass such as the massive vector boson, 
its propagator ${G_{\mu \mu}}^{\rm off}(r)$ would be described by the 
Yukawa-type function $\sim \exp(-M_{\rm off}r)/r^{3/2}$, and if 
we find the linear behavior for $\ln (r^{3/2} {G_{\mu \mu}}^{\rm off}(r))$, 
the mass $M_{\rm off}$ can be extracted from its slope.
As a result, we find that the off-diagonal gluon has the large mass 
$M_{\rm off} \simeq 1$ GeV as shown in Fig. 1.
That is to say, the interaction range of the off-diagonal gluon 
is limited within the short distance corresponding to its inverse 
mass $M_{\rm off}^{-1}\simeq 0.2$ fm.
Thus, the off-diagonal gluon does not contribute to the long-range physics,  
which predicts general infrared abelian dominance in the MA gauge.

\par
As for monopole condensation, we study the inter-monopole potential
and evaluate the dual gluon mass using the SU(2) lattice QCD 
simulation\cite{atanaka}.
The dual gluon field $B_{\mu}$ is introduced to satisfy 
$\partial_{\mu} B_{\nu}-\partial_{\nu} B_{\mu}={}^*F_{\mu \nu}$
and $\partial_{\mu} {}^*F_{\mu \nu}=k_{\nu}$. Here, $k_{\nu}$ is the 
color-magnetic monopole current.
The idea used here is quite similar to the evaluation 
of the off-diagonal gluon mass.
If monopole condensation is occurred, the dual gluon 
becomes massive due to the dual Higgs mechanism.
Then, its mass $m_B$ can be extracted by fitting the 
Yukawa potential $V_{\rm M}(r) \sim -\exp(-m_B r)/r$, 
since the dual gluon behaves as 
the massive vector boson.
From this analysis, we find that the dual gluon 
acquires the mass $m_B \simeq 0.5$ GeV as shown in Fig. 2, 
which is just the quantitative evidence of 
monopole condensation.

\par
As an interesting application of abelian dominance for the 
inter-quark potential,
we can calculate the quark single-particle potential $U(x)$
for the low-lying hadron ($m_q$=300 MeV). 
Here, $U(x)$ is defined by the superposition of the inter-quark potential 
$V(r)=-c/r +\sigma r$ ($\sigma \simeq 1$ GeV/fm, $c \simeq 0.4$)
with the weight of the color charge distribution 
$\rho({\bf x})=\bar{\psi}_q\gamma_0 \vec{H} \psi_q \cdot \vec{Q}$
as $\vec{Q}^2 U(x)=\int d^3 x \rho ({\bf x'})V(|{\bf x-x'}|)$.
Solving the self-consistent equations between the quark 
wave function and the quark potential,
we obtain the color charge distribution 
and the quark single-particle potential as shown in Figs.~3 and 4.
The color charge distribution is spread over a intermediate 
region $ r \sim 0.5$ fm. 
The quark single-particle potential is found to be 
flat at the short distance, which can be connected with the bag model.

\par
\noindent
\vspace{-3cm}
\begin{figure}[t]
\hspace{1cm}
\begin{minipage}[hbt]{6cm}
\epsfxsize=6cm
\vspace{-0.6cm}
\epsfbox{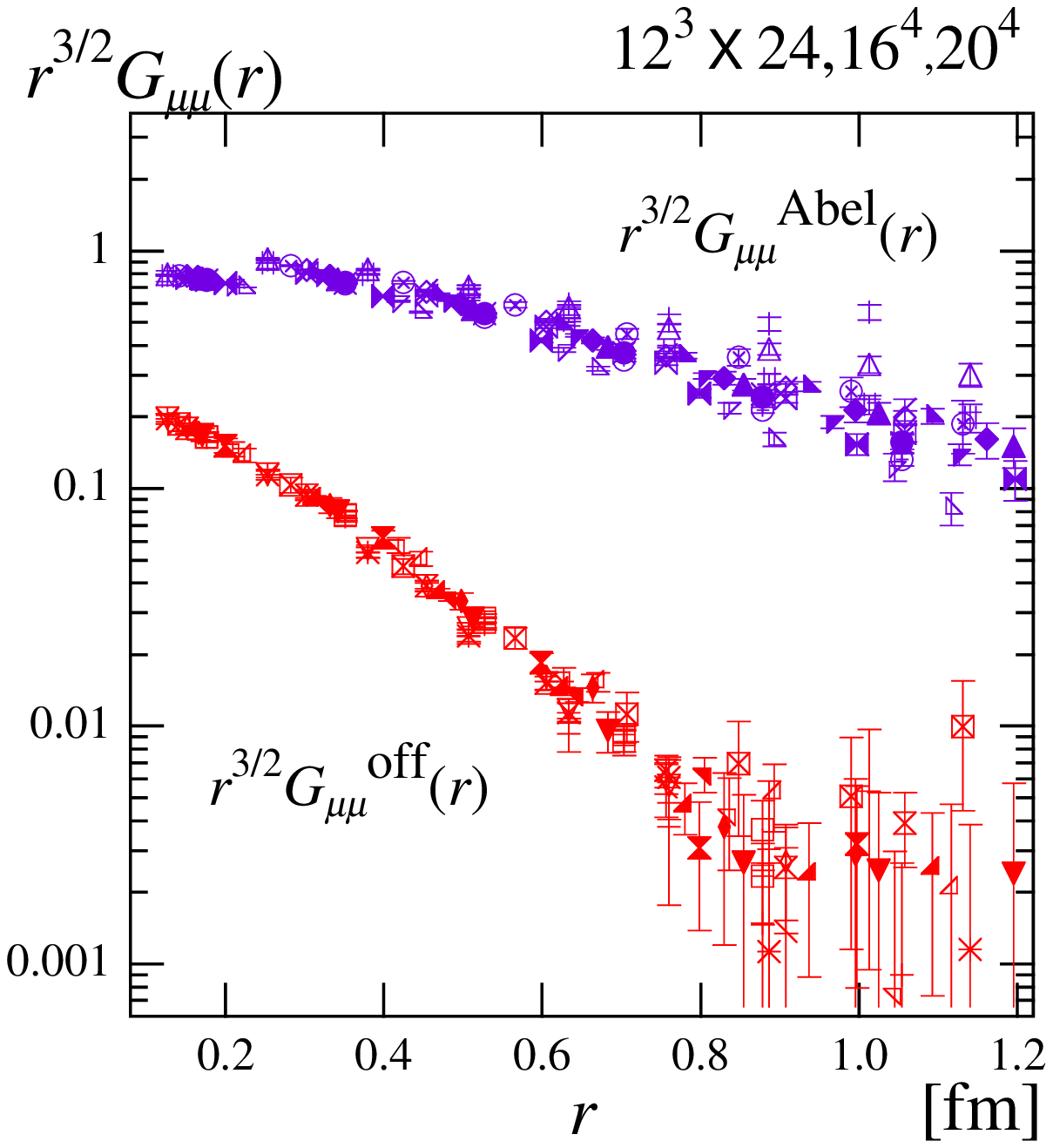}
\vspace{-0.1cm}
Fig.~1. The gluon propagator 
as a function of 4-dimensional distance $r$ in the MA gauge.
\end{minipage}

\vspace{-6.7cm}
\hspace{8cm}
\begin{minipage}[t]{7.5cm}
\epsfxsize=7.5cm
\epsfbox{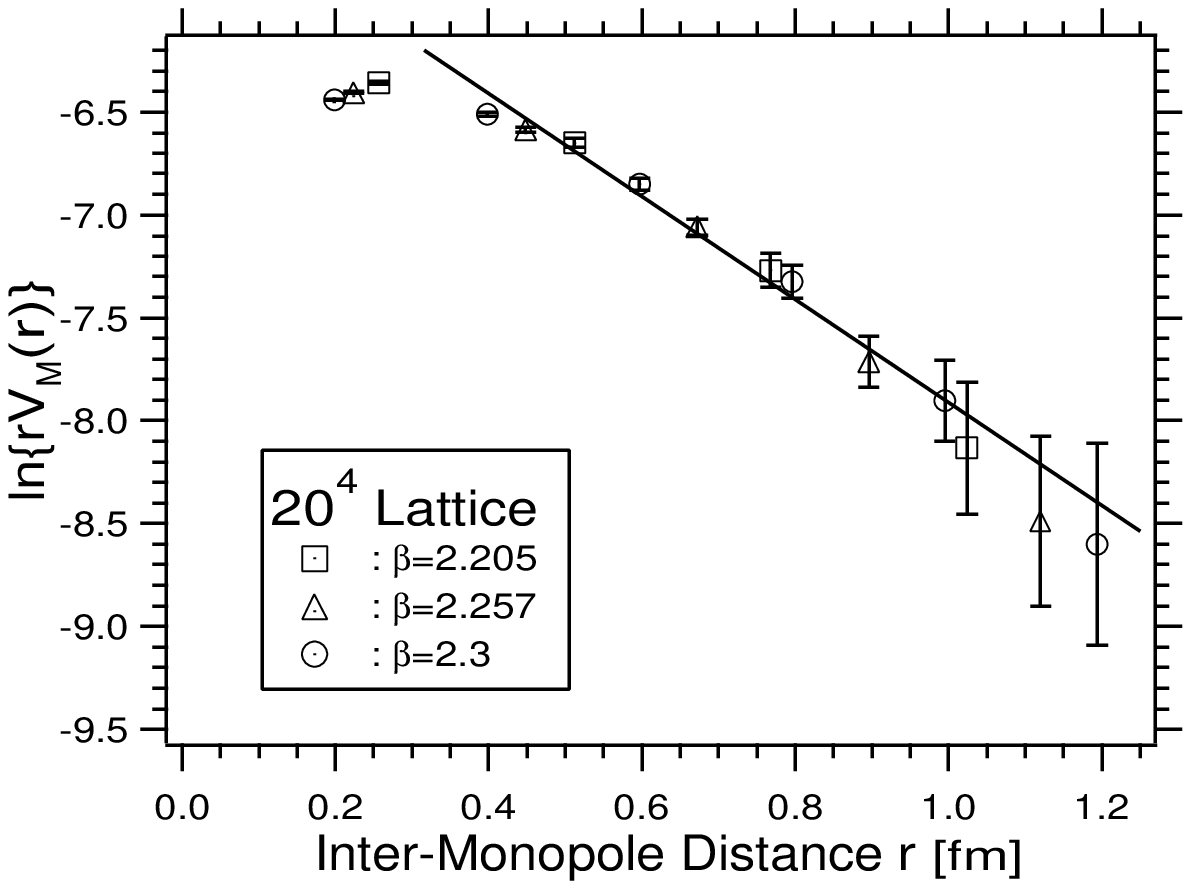}
\vspace{-0.1cm}
Fig.~2. The inter-monopole potential 
as a function of 4-dimensional distance $r$ in the MA gauge.
\end{minipage}
\vspace{-0.84cm}
\end{figure}


\par
\noindent
\begin{figure}[t]
\hspace{1cm}
\begin{minipage}[hbt]{6.5cm}
\vspace{0.5cm}
\epsfxsize=6cm
\vspace{-0.9cm}
\epsfbox{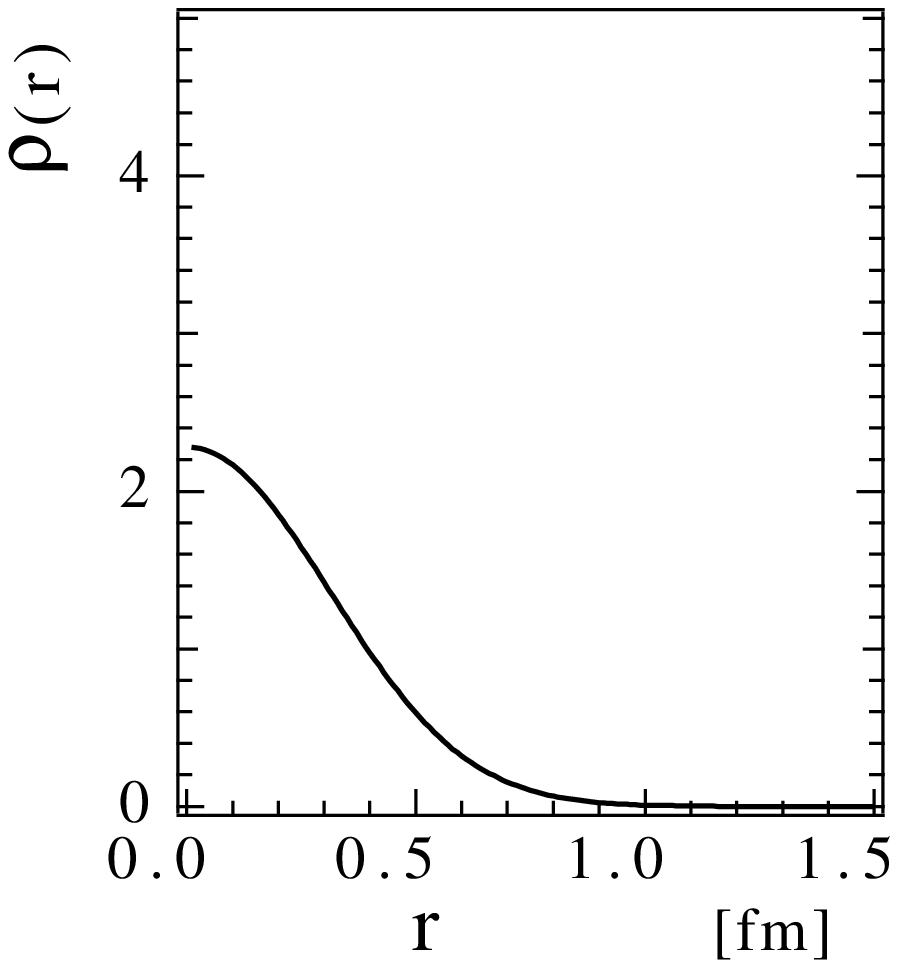}
\vspace{-0.3cm}
Fig.~3. The color charge distribution for the low-lying hadron.
\end{minipage}

\vspace{-7.4cm}
\hspace{8.5cm}
\begin{minipage}[t]{6.5cm}
\epsfxsize=6.2cm
\epsfbox{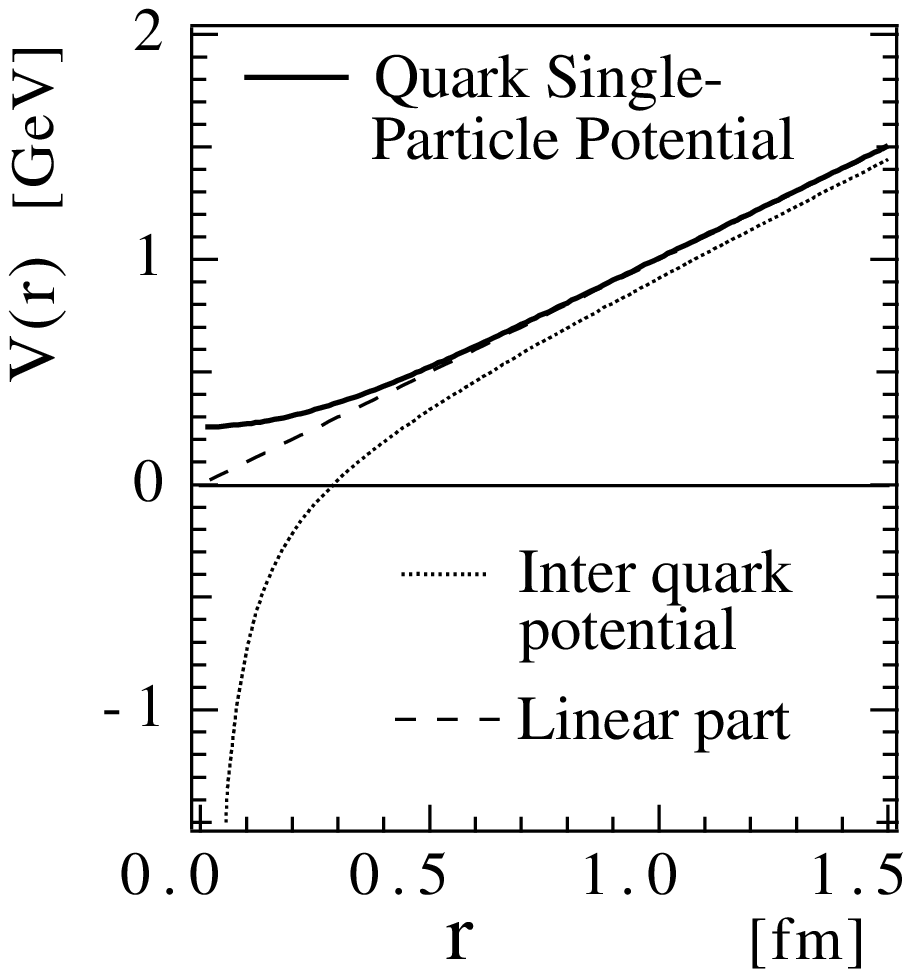}
\vspace{-0.3cm}
Fig.~4. The quark single particle potential in the 
low-lying hadrons.
\end{minipage}
\vspace{-.4cm}
\end{figure}

\newpage
\section{The DGL theory and application to the scalar glueball analysis}

\par
The DGL theory can be constructed by taking into account 
abelian dominance and monopole condensation in the MA gauge in QCD.
The DGL lagrangian is given as
\vspace{-0.1cm}
\begin{equation}
{\cal L}_{\rm DGL} \!\!=\!\!
-\frac{1}{4} \! \left (\partial_{\mu}\vec{B}_{\nu} 
\!\! - \! \partial
_{\nu}\vec{B}_{\mu} 
\!-\!
\frac{1}{n \cdot \partial}\varepsilon_{\mu \nu \alpha \beta}
n^{\alpha} \vec{j}^{\beta} \right )^2
\!\!
+\sum_{\alpha=1}^3 \left [ \left | \left (\partial_{\mu}
\! + \! ig\vec{\epsilon}
_{\alpha}{\cdot}\vec{B}_{\mu} \right )
\! \chi_{\alpha} \right |^2
\!\! - \! 
\lambda \left ( \left |\chi_{\alpha} \right |^2
\!\! - \! v^2 \right )^2 \right ],
\end{equation}
\vspace{-0.1cm}
where $\vec{B}_{\mu}$ and $\chi_{\alpha}$ denote the dual gluon field with 
two components $(B_{\mu}^3, B_{\mu}^8)$ and the complex scalar monopole 
field, respectively. The quark field is included in the current
$\vec{j}_{\mu} = e \bar{q}\gamma_{\mu} \vec{H} q$.
Here, $\vec{\epsilon}_a$ is the root vector of SU(3) algebra, 
and $n^{\mu}$ denotes an arbitrary constant 4-vector, which corresponds
to the direction of the Dirac string.
The gauge coupling $e$ and the dual gauge coupling $g$ 
hold the relation $eg=4\pi$.

\par
Monopole condensation is characterized by 
$\langle 0|\chi_{\alpha}|0 \rangle$=$v$, and 
the dual gluon field acquires the mass $m_B=\sqrt{3}gv \simeq$ 0.5 GeV
through the dual Higgs mechanism.
Then, the DGL theory describes the QCD vacuum as the dual superconductor.
The quark confinement is explained by the dual Meissner effect, 
which forces the color-electric field between the quarks to form the 
flux-tube configuration, and leads the linear inter-quark potential.
This flux-tube also provides intuitive picture of hadrons.
If we apply this flux-tube picture to the glueball, it would
be identified with the flux-tube ring, since the glueball is considered
to have no valence quarks, and the lowest state is 
the scalar glueball.
From the flux-tube ring solution in the DGL theory, we find the 
mass and the size of the scalar glueball
as 1.6 GeV and 0.5 fm, respectively\cite{koma}.
It is interesting to note that this mass spectrum is consistent with the
recent lattice QCD results $M(0^{++})$ = 1.50 - 1.75 GeV\cite{lattice-gb}.

\par
Here, we find another aspect of the scalar glueball in the DGL theory, 
which is closely related to the dual Higgs mechanism.
Taking monopole condensation into account, the monopole field 
can be defined as 
$\chi_{\alpha} \equiv \left (v + \tilde \chi_{\alpha}/\sqrt{2} 
\right ) e^{i \eta_{\alpha} / v}$,
where $\tilde \chi_{\alpha}$ and $\eta_{\alpha}$ are real variables
denoting the magnitude of the vacuum fluctuation and the phase, respectively.
Here, $\alpha$=1, 2, 3 labels the color-magnetic charge of the monopole field,
dual-red, dual-blue and dual-green.
Since the origin of the monopole field is the off-diagonal gluon 
field in the MA gauge in QCD, this field 
$\tilde \chi_{\alpha}$ would present the scalar gluonic excitation
corresponding 

\newpage
\noindent
to the dual Higgs particle.
In particular, the Weyl symmetric monopole field defined by  
$\tilde \chi^{(0)} \equiv (\tilde \chi_1 
+\tilde \chi_2 +\tilde \chi_3)/\sqrt{3}$ 
is the color-singlet field\cite{ichie} so that it can be regarded as 
the scalar glueball with the mass $m_\chi=2\sqrt{\lambda}v \simeq 1.6$ GeV.
Although the relation between the flux-tube ring is not clear,
it can be considered as another feature of the scalar glueball.

\par
Here, we concentrate on the calculation of the 
$\tilde \chi^{(0)} q \bar{q}$ vertex function, which plays an 
important role to understand how the scalar glueball interacts 
with the quarks.
The lowest diagram is shown in Fig. 5. The scalar glueball interacts
with the dual gluon at first, and then, the dual gluon interacts with 
the quarks.
We show the typical behavior of the vertex function in the scalar channel, 
as a function of the coupled quark momentum in Fig. 6. 
Here, we have set $p\cdot q$=0 for simplicity.
We find that the heavy quark ($m_c \simeq 1.6$ GeV) interacts with the scalar 
glueball about four times stronger than the light 
quarks ($m_{u,d,s}\simeq 0.3-0.5$ GeV).
It seems to indicate the flavor dependence of the interaction of the 
scalar glueball.
It is interesting to study how this interaction property 
reflects on the scalar glueball decay into the two pseudo-scalar mesons
and the glueball-quarkonium mixing states, which are now investigating.

\begin{figure}[hbt]
\vspace{-0.5cm}
\epsfxsize=6cm
\vspace{0.3cm}
\hspace{0.5cm}
\epsfbox{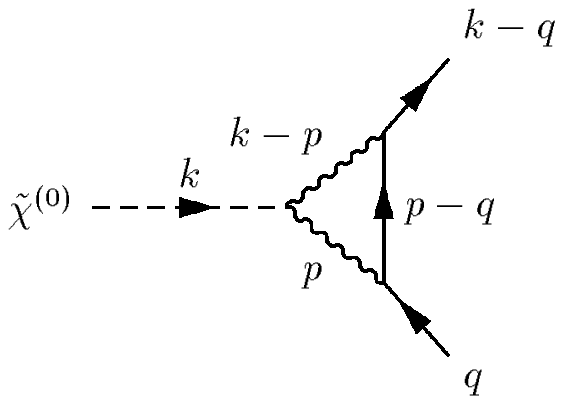}
\end{figure}
\vspace{-0.7cm}
\begin{minipage}[hbt]{6.0cm}
Fig.~5(upper). $\tilde \chi^{(0)}q\bar{q}$ vertex.\\

\vspace{-0.3cm}
\noindent
Fig.~6(right). The vertex function of $\tilde \chi^{(0)}q\bar{q}$ 
in the scalar channel vs. coupled quark momentum.
\end{minipage}

\begin{figure}[hbt]
\vspace{-1.3cm}
\epsfxsize=7.5cm
\vspace{-7cm}
\hspace*{7.5cm}
\epsfbox{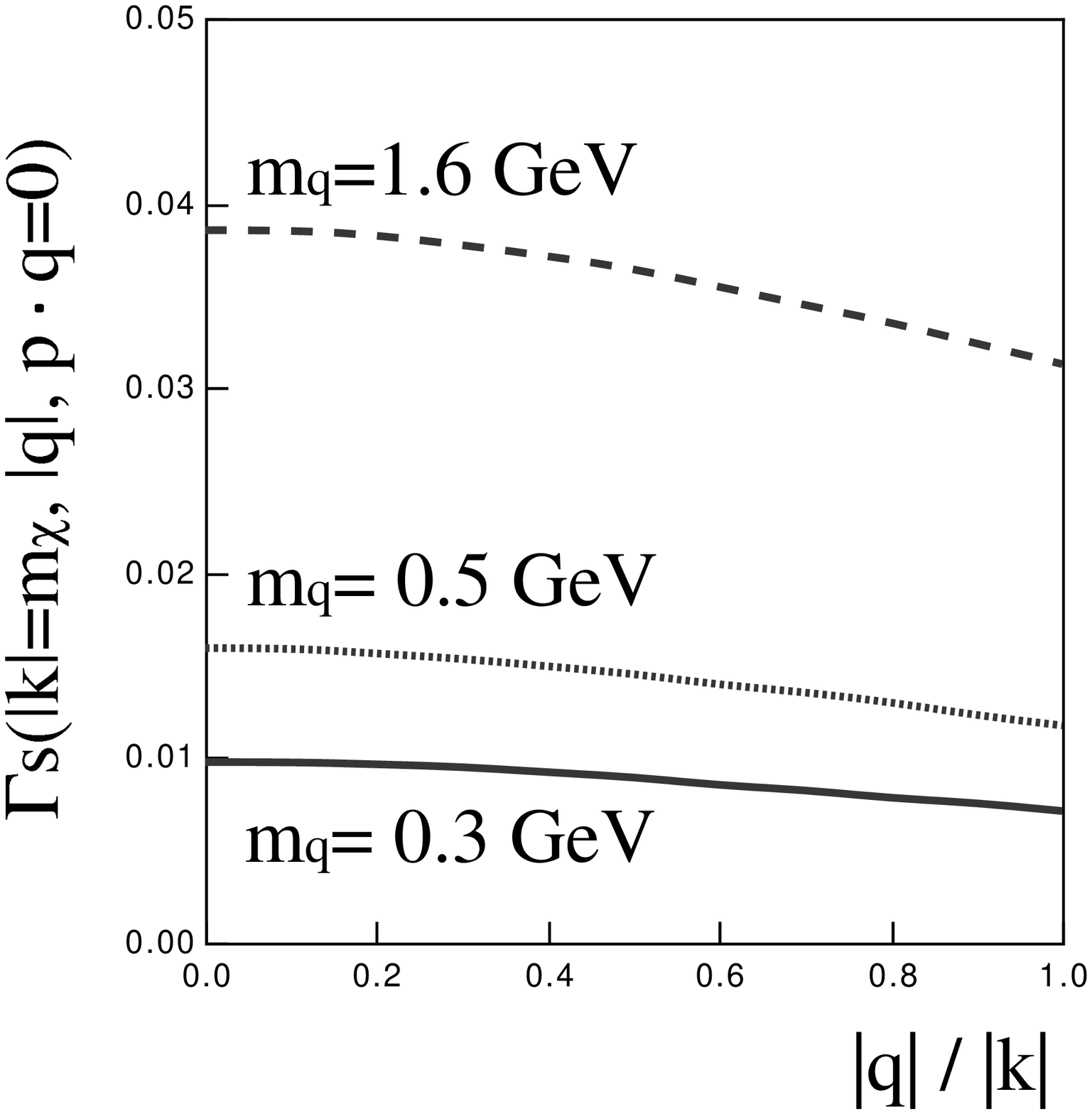}
\end{figure}

\end{document}